\documentclass[%
 reprint,
nofootinbib,
 amsmath,amssymb,
 aps,
]{revtex4-2}

\usepackage{graphicx}
\usepackage{dcolumn}

\usepackage{comment}
\DeclareUnicodeCharacter{2212}{-}

\newcommand{\cO}{\mathcal{O}}
\newcommand{\cR}{\mathcal{R}}
\newcommand{\cS}{\mathcal{S}}
\newcommand{\cT}{\mathcal{T}}

\newcommand{\p}{\partial}

\newcommand{\gt}{\tilde{g}}
\newcommand{\ut}{\tilde{u}}

\newcommand{\be}{\begin{equation}}
\newcommand{\ee}{\end{equation}}

\begin{document}

\title{On harvesting physical predictions \\ from asymptotically safe quantum field theories}%
 
\author{Frank Saueressig}
\email{f.saueressig@science.ru.nl}
\author{Agustín Silva}
\email{agustin.silva@ru.nl}
\affiliation{High Energy Physics Department, Institute for Mathematics, Astrophysics, and Particle Physics, Radboud University, Nijmegen, The Netherlands.}

\date{\today}

\begin{abstract}
Asymptotic safety is a powerful mechanism for obtaining a consistent and predictive quantum field theory beyond the realm of perturbation theory. It hinges on an interacting fixed point of the Wilsonian renormalization group flow which controls the microscopic dynamics. Connecting the fixed point to observations requires constructing the set of effective actions compatible with this microscopic dynamics. Technically, this information is stored in the UV-critical surface of the fixed point. In this work, we describe a novel approach for extracting this information based on analytical and pseudo-spectral methods. Our construction is illustrated at the level of the three-dimensional Ising model and easily generalizes to any asymptotically safe quantum field theory. It also constitutes an important step towards setting up a well-founded swampland program within the gravitational asymptotic safety program.
\end{abstract}

\maketitle

\section{Introduction}
The Wilsonian renormalization group (RG) \cite{Wilson:1973jj} provides a powerful tool for understanding the impact of statistical or quantum fluctuations in a given physical system \cite{berges2002non,Dupuis:2020fhh,Reuter:2012id,zinn2021quantum}. It organizes fluctuations in momentum shells and integrates out the modes, starting from the most energetic ones and subsequently moving to lower energies. This procedure creates RG trajectories which connect effective descriptions of the same system at different values of the coarse-graining scale.  

A major success of this approach is a comprehensive picture of critical phenomena and universality which are readily explained in terms of RG fixed points controlling the theory's infrared (IR) behavior. Typically, one has to tune a small number of parameters so that a RG trajectory is dragged into the fixed point in the limit where all fluctuations are integrated out. These trajectories then ``forget'' their microscopic origin and physical quantities like correlation functions are completely dictated by the properties of the IR fixed point. 

From a physics perspective, one may also be interested in situations where a RG fixed point provides the microscopic description of the system. This situation applies to asymptotically free quantum field theories like quantum chromodynamics where the relevant fixed point is the free theory. Asymptotic safety, first proposed in \cite{weinberg1978critical,weinberg1979ultraviolet}, generalizes this construction to interacting fixed points, also called non-Gaussian fixed points (NGFPs). While there are situations (e.g.\ in the context of gravity) where it is unlikely that the scaling regime associated with the fixed point can be probed directly by experiments due to the required energy scales, it nevertheless equips the construction with predictive power. By definition, RG trajectories whose ultraviolet (UV) completion is provided by the fixed point span its UV-critical surface $\mathcal{S}_{\rm UV}$. In the vicinity of the fixed point, this surface can be found by linearizing the RG flow and identifying the UV-attractive directions. 
Typically, $\mathcal{S}_{\rm UV}$ is finite-dimensional and embedded in a larger space called the theory space $\cT$. This allows to predict values for renormalized couplings based on the free parameters specifying a RG trajectory within $\mathcal{S}_{\rm UV}$. These predictions are obtained from the end-point of the RG trajectory where all fluctuations have been accounted for. In the case of an IR fixed point, this limit is controlled by the fixed point itself, so that its properties are directly related to observables. For an UV fixed point, this link is highly non-trival though and requires the construction of $\cS_{\rm UV}$. Depending on the dimensions of $\cS_{\rm UV}$ and $\cT$ this becomes  cumbersome rather quickly. A prototypical example is provided by asymptotically safe quantum gravity supplemented by the matter degrees of freedom of the standard model, where dim($\cS_{\rm UV}$) is expected to be of order twenty \cite{Pastor-Gutierrez:2022nki}. This surface should then be embedded in an even larger space also comprising couplings associated with beyond the standard model physics for making predictions. It is clear that the canonical approach of mapping out $\cS_{\rm UV}$ based on a shooting-method is intractable in such cases. Nevertheless, a solid knowledge about the effective actions within $\cS_{\rm UV}$ is crucial for making predictions and potentially falsifying the asymptotic safety hypothesis.

The goal of this letter is to introduce novel strategies which allow to map out $\cS_{\rm UV}$ in a computationally efficient way using pseudo-spectral methods \cite{boyd2013chebyshev,shen2011spectral} which are readily boosted by machine-learning algorithms to improve convergence. We illustrate the generic algorithm based on the RG flow of a scalar field theory in $d=3$ which has already been extensively studied, e.g., in \cite{Tetradis:1993ts,Morris:1994ki,Adams:1995cv,Litim:2002cf,Canet:2003qd,Juttner:2017cpr,DePolsi:2020pjk,Balog:2019rrg}. Owed to the flexibility of the RG \cite{Dupuis:2020fhh}, our algorithm is applicable to a much wider range of settings, including the study of phase transitions in statistical physics \cite{berges1997equation}, estimates of the rate of spontaneous nucleation \cite{strumia1999region}, strongly interacting and high temperature theories \cite{tetradis1994high}, up to quantum gravity building on the gravitational asymptotic safety program \cite{percacci2017introduction,reuter2019quantum}. 
\bigskip

\section{RG flows, fixed points \\ and free parameters}
A primary tool for computing Wilsonian RG flows is the Wetterich equation 
\cite{wetterich1993exact} (also see \cite{,Ellwanger:1993mw,morris1994exact,Bonini:1992vh} for related early works and \cite{Reuter:1993kw, Reuter:1996cp} for its generalization to gauge theories and gravity, respectively),\footnote{Our construction is readily adapted to other RG equations like the Polchinski equation \cite{polchinski1984renormalization}. Essentially, it applies to any situation where a vector field is used to generate a constraint surface also including the case of IR-repulsive surfaces of IR-fixed points.}
\begin{equation}\label{FRGE}
    k \p_k \Gamma_k =\frac{1}{2}\mathbf{Tr}\left[ 
    \left( \Gamma_k^{(2)} + \cR_k \right)^{-1} k \p_k \cR_k \right],
\end{equation}
which encodes the dependence of the effective average action $\Gamma_k$ on the coarse-graining scale $k$, see \cite{Pawlowski:2005xe,Rosten:2010vm,Saueressig:2023irs} for more details. The effective average action lives on theory space $\cT$. By definition, $\cT$ contains all action functionals which can be constructed from the field content of the theory and are compatible with its symmetry requirements. For the purpose of this work, we are interested in complete (approximate) solutions of \eqref{FRGE}, $k \mapsto \Gamma_k$, which interpolate between a NGFP in the limit $k\rightarrow \infty$ (asymptotic safety) and the effective action $\Gamma \equiv \lim_{k \rightarrow 0} \Gamma_k$ associated with observables.

Introducing a basis $\{\cO_n\}$ on $\cT$, $\Gamma_k$ can be expanded as 
\be\label{Gammaexpansion}
\Gamma_k = \sum_{n} \; \tilde{g}^n_k \, \cO_n \, .
\ee
Here the $\tilde{g}^n_k$ are the dimensionful couplings of the theory which depend on $k$. Denoting the classical mass-dimension of $\tilde{g}^n_k$ by $d_n$, these couplings are conveniently traded for their dimensionless counterparts 
\be\label{dimless}
g^n_k \equiv \gt^n_k \, k^{-d_n} \, . 
\ee
In explicit computations, one has to resort to approximations of $\Gamma_k$ by restricting the set of operators $\{\cO_n \}$ retained in eq.\ \eqref{Gammaexpansion}. These truncations can either be functional where one tracks the $k$-dependence of field dependent functions by solving partial differential equations or, more drastically, limited to a finite number of operators, $N$, only. The second setting naturally connects to an effective field theory description of the dynamics, and we will work in this setting in the sequel. The couplings $g^n$, $n=1,\cdots,N$ can then be read as coordinates on $\mathbb{R}^N$. Substituting \eqref{Gammaexpansion} into \eqref{FRGE}, the beta functions capturing the $k$-dependence of $g_k^n$ can be read off as the coefficients multiplying the basis elements $\{\cO_n\}$:
\be\label{betadef}
k \p_k g_k^n = \beta^n(g^m) \, . 
\ee
In the sequel, we will assume that the beta functions  have been computed and admit a NGFP suitable for asymptotic safety.\footnote{Establishing the existence of a NGFP based on the functional renormalization group is in itself highly non-trivial and may require going beyond finite-dimensional truncations in order to establish that the finite-dimensional expansion converges against a functional with physically admissible boundary conditions. This is a research line in itself and will not be covered in this work.} 

By definition, fixed points $\{g^n_*\}$ satisfy $\beta^n(g^m_*) = 0, \forall n$. The dimension of $\cS_{\rm UV}$ associated with the fixed point is readily inferred by linearizing the beta functions
\be\label{betalin}
k \p_k g^n_k \simeq \sum_m B^n{}_m (g^m_k - g^m_*) \, , \quad B_m^{\,\,\,\,\,n} \equiv \left. \frac{\p \beta^n}{\p g^m} \right|_{g = g_*} \, . 
\ee
These equations are readily solved in terms of the stability coefficients $\theta_I$ and right-eigenvectors $V_I$, satisfying $B V_I = - \theta_I V_I$. Separating the stability coefficients $\{\theta_I\} \rightarrow \{ \theta_\alpha , \tilde{\theta}_\mu \}$ where $\alpha = 1,\cdots,\dim(\cS_{\rm UV})$ and $\mu = 1,\cdots,N-\dim(\cS_{\rm UV})$ enumerate the stability coefficients with Re$(\theta_\alpha) > 0$ and Re$(\tilde{\theta}_\mu) < 0$, respectively, one has
\be\label{linsol}
g^n_k \approx g^n_* + \sum_\alpha C_\alpha \, V_\alpha^n \left( \frac{k_0}{k} \right)^{\theta_\alpha } + \sum_\mu \tilde{C}_\mu \, \tilde{V}_\mu^n \left( \frac{k_0}{k} \right)^{\tilde{\theta}_\mu } \, . 
\ee
Here $k_0$ is a reference scale and $C_\alpha, \tilde{C}_\mu$ are integration constants labelling the specific solutions. The eigendirections $V_\alpha$ $(\tilde{V}_\mu)$ are attractive (repulsive) in the limit $k \rightarrow \infty$. Thus asymptotic safety fixes $\tilde{C}_\mu = 0$ while the ${C}_\alpha$ are free parameters. RG trajectories where $\tilde{C}_\mu = 0$ span 
$\cS_{\rm UV}$. 

\begin{figure}[t]
     \centering
     \includegraphics[width=\columnwidth]{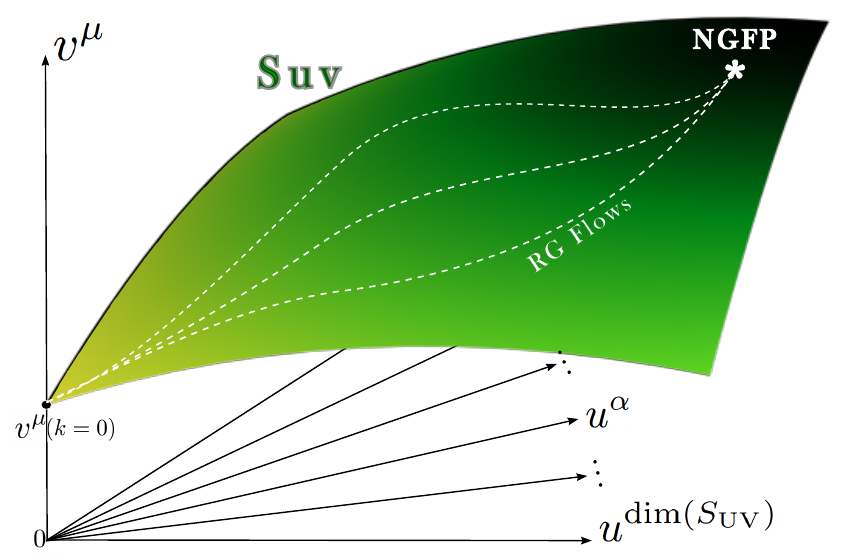}
     \caption{Using the choice of couplings introduced below eq.\ \eqref{limits}, the $k \to 0$ limit of the  couplings $u^\alpha$ is $\{u^\alpha\} = 0$. Every complete RG trajectory emanating from the NGFP gives rise to an effective action whose parameters $v^\mu$ are read off from the intersect of $\mathcal{S}_{\rm UV}$ with the axes $\{u^\alpha\} = 0$.}
     \label{fig:CHTh}
 \end{figure}

Subsequently, we reformulate the conditions on the integration constants in terms of the dimensionless couplings. For this purpose, we split
\be\label{embedding1}
g^n \mapsto \{ u^\alpha \, , \, v^\mu \} \, , \quad
\beta^n \mapsto \{ \beta^\alpha \, , \, \beta^\mu \} \, . 
\ee
The basic idea illustrated in Fig.\ \ref{fig:CHTh} is to use the $u^\alpha$ as coordinates on $\cS_{\rm UV}$.\footnote{Depending on the structure of $\cS_{\rm UV}$, the initial coordinate system may not cover the entire surface. In this case, $\cS_{\rm UV}$ has to be covered with multiple coordinate patches. A simple illustration is provided by the surface $x^2+y^2-1=0$, embedded into $\mathbb{R}^2$, where there is no single-valued function $y(x)$ that covers the entire surface.  In the context of the RG such coordinate changes have recently been discussed in \cite{Buccio:2022egr}.} Its embedding into $\mathbb{R}^N$ is then given by a set of generating functions
\be\label{surface1}
F_\nu(u, v) = 0 \, , \qquad \nu = 1,\cdots,N-\dim(\cS_{UV}) \, . 
\ee
 The generating functions are combinations of couplings whose values are conserved along the RG flow. We adopt the convention that $\cS_{\rm UV}$ is generated by setting the conserved quantity to zero. Solving these relations for $v^\mu$ then determines the values of the couplings $v^\mu$ in terms of the coordinates on the surface
\be\label{prediction}
\left. v^\mu \right|_{S_{\rm UV}} = v^\mu(u^\alpha) \, . 
\ee

In the vicinity of the NGFP, the functions $F_\nu$ can be found as follows. The vectors $\{V_\alpha, \tilde{V}_\mu\}$ span the tangent space at $\{g_*^n\}$. The vectors $V_\alpha$ are tangent to $\mathcal{S}_{\rm UV}$ while, in general, the $\tilde{V}_\mu$ are not normal to the surface. A set of normal vectors can then be constructed in two steps. First, we project the $V_\alpha$ into a orthogonal basis $t_\alpha$, carrying out a Gram-Schmidt process. Starting from the vectors $\tilde{V}_\mu$, the set of normals is obtained as 
\be
n_\mu = \tilde{V}_\mu - \sum_\alpha \, t_\alpha \frac{\langle t_\alpha | \tilde{V}_\mu \rangle}{\langle t_\alpha | t_\alpha \rangle} \, ,
\ee
where $\langle x|y \rangle$ is the standard vector product on $\mathbb{R}^N$. At the linearized level, the constrained equations \eqref{surface1} are then given by
\be\label{surface2}
F_\mu^\ast(g^n) = \sum_n \, n^n_\mu \, \left( g^n - g^n_* \right) \, . 
\ee
Here $\ast$ emphasizes that this is the expansion of the full $F_\mu$ to linear order at the NGFP. One may then single out couplings $u^\alpha$ according to \eqref{embedding1} and solve these linear relations to express the $v^\mu$ in terms of these parameters.
\section{Constructing $\cS_{\rm UV}$}
\label{sect.3}
Finding the IR-endpoints of asymptotically safe RG trajectories requires the construction of $\cS_{\rm UV}$ beyond the linear approximation \eqref{surface2}. Technically, it is then convenient to work with a compact subregion of $\mathbb{R}^N$ which contains the NGFP and the endpoints. In order to achieve this, it is instructive to discuss the $k\rightarrow 0$-limit of eq.\ \eqref{dimless}. Keeping $\lim_{k \rightarrow 0}\gt^n_k$ fixed and finite, which corresponds to the end-point of the RG trajectory being in the classical regime,\footnote{One may also be interested in situations where the $k \rightarrow 0$-limit is provided by a NGFP. In this case, our construction needs to be modified taking the anomalous scaling dimensions of the couplings into account.} the dimensionless couplings with a non-zero canonical mass-dimension $d_n$ either approach zero or diverge
\be\label{limits}
\begin{array}{ll}
\lim_{k \rightarrow 0} g^n_k =  0 \, , \qquad & d_n < 0 \, , \\[1.2ex] 
\lim_{k \rightarrow 0} g^n_k =  \pm \infty \, , \qquad & d_n > 0 \, .
\end{array}
\ee
Couplings with $d_n=0$ take the finite value of the coupling appearing in the effective action. Based on these insights, we choose the couplings $\ut^\alpha$ to have a negative mass dimension. This can always be achieved by making the identification $\ut^n = 1/\gt^n$ if the initial coupling comes with a positive mass dimension. This has the advantage that IR-endpoints in a classical regime are located at $\{u^\mu\} = 0$. Furthermore, we can take the $v^\mu$ to be dimensionless by constructing appropriate ratios of the couplings, see \eqref{Upot} for an explicit example. The physical predictions for the couplings $v^\mu$ are then read off at $\{ u^\alpha \}= 0$.  

\subsection{Master equation and explicit form}
We now derive our master equation encoding the structure of $\cS_{\rm UV}$. Taking the $k$-derivative of the generating functions $F_\mu(u,v)$ and using the definition of the beta functions \eqref{betadef} leads to
\be\label{master1}
\sum_{\alpha} \, \frac{\p F_\mu}{\p u^\alpha} \, \beta^\alpha + 
\sum_\nu \frac{\p F_\mu}{\p v^\nu} \, \beta^\nu = 0 \, . 
\ee
At this point we assume that $\det(\frac{\p F_\mu}{\p v^\nu}) \not = 0$, so that the implicit function theorem guarantees that the system $F_\mu = 0$ has a (local) solution in the form \eqref{prediction}. 

The master equation can be turned into an system of non-linear partial differential equations for the functions \eqref{prediction}. Substituting $F_\mu(u,v) = v^\mu - v^\mu(u)|_{\cS_{\rm UV}}$, which obviously satisfies $F_\mu|_{\cS_{\rm UV}} = 0$, leads to the linear partial differential equation
\be\label{master2}
\sum_{\alpha} \, \beta^\alpha \, \frac{\p v^\mu}{\p u^\alpha} = \beta^\mu \, . 
\ee
The boundary conditions for the system are given in the linearized regime, eq.\ \eqref{surface2}.  Solutions can then be obtained either by solving the system \eqref{master1} or \eqref{master2} using numerical methods or extending the power series \eqref{surface2} to include higher-order terms. The latter strategy requires checking whether the endpoint of interest is within the radius of convergence of the series.

\subsection{Implicit solutions}
The key strength of the master equation \eqref{master1} is that it is a multi-linear, first order system of partial differential equations for the generating functions $F_\mu(u,v)$. Finding the implicit form of $\cS_{\rm UV}$ by solving this system leads to a significant reduction in complexity as compared to solving the explicit system \eqref{master2}. In particular, the linear nature of the master equation makes the use of pseudo-spectral methods and collocation techniques \cite{fornberg1994review, hussaini1983spectral} a highly efficient tool for obtaining solutions.\footnote{For earlier applications of pseudo-spectral methods in the context of the functional renormalization group see \cite{Borchardt:2016pif,Borchardt:2015rxa,Draper:2020bop,Fehre:2021eob}.} 

The basic idea is to select a dense set of functions $\{ \psi_i(u,v) \}$. Typical examples used in practice include Chebyshev polynomials \cite{boyd2013chebyshev} or radial basis functions \cite{buhmann2000radial}. The generating functions are then approximated by expanding in this basis
\be\label{spectral-expansion}
F_\mu(u,v) \simeq \sum_{i=1}^{N_p} p_{\mu,i} \, \psi_i(u,v) \, , 
\ee
i.e., the functions are determined by the $N_p \times (N-\dim(\cS_{\rm UV}))$ free parameters $p_{\mu,i}$. Since \eqref{master1} contains derivatives of $F_\mu(u,v)$ only, it determines solutions up to an additive constant. Following \eqref{surface2}, we fix this freedom by demanding that
\be\label{boundary1}
F_\mu(u_*,v_*) = 0 \, . 
\ee
Secondly, the first derivatives of $F_\mu$ evaluated at the fixed point are taken to agree with the linearized approximation
\be\label{boundary2}
\left. \frac{\p F_\mu}{\p u^\alpha} \right|_{g^n_*} = \frac{\p F^\ast_\mu}{\p u^\alpha} \, , \qquad
\left. \frac{\p F_\mu}{\p v^\nu} \right|_{g^n_*} = \frac{\p F^\ast_\mu}{\p v^\nu} \, .  
\ee
This enforces that some of the expansion coefficients in \eqref{spectral-expansion} must take non-vanishing values thereby eliminating the trivial solution. 

Subsequently, one chooses $N_p$ collocation points.  It is natural to take the NGFP as one of these points and construct a grid covering the parameter space of interest. Substituting the expansion \eqref{spectral-expansion} into \eqref{master1} and evaluating the resulting expressions at the collocation points then leads to a system of algebraic equations which determines the coefficients 
$p_{\mu,i}$. The predictions $v^\mu$ compatible with asymptotic safety are then found by evaluating the approximate solution at $\{ u^\alpha \}= 0$.

\emph{Systematic improvements.} In the simplest case, one may arrange the collocation points $(u^\alpha_{c,i}, v^\mu_{c,i})$ in a regular lattice and expand the generating functions in terms of Multivariate Cauchy Distributions (MCDs) 
\begin{equation}
    \psi_{i}(u,v)= \left(1 +  \sum_\alpha \frac{(u^\alpha - u^\alpha_{c,i})^2}{\sigma^2} + \sum_\mu \frac{(v^\mu - v^\mu_{c,i})^2}{\sigma^2}  \right)^{-1} \, . 
\end{equation}
Here $\sigma$ is a smoothness parameter whose typical value is chosen to be of the order of the average separation between the collocation points. Since the substitution of \eqref{spectral-expansion} into the system \eqref{master1} results in a linear equation, taking one collocation point for each MCD leads to a unique solution for the parameters $p_{\mu,i}$. 

Depending on the concrete application, this ``basic'' setup may be improved by working with less parameters than collocation points, and/or with base functions more suitable for describing sharp edges, periodicity, or optimized convergence \cite{pachon2009barycentric}. Moreover, the positioning of the collocation points may be optimized near interesting regions. In the case of using less parameters than collocation points, the construction is overdetermined and finding solutions turns into an optimization problem, where one could use machine learning \cite{powell1981approximation} techniques, such as gradient descent methods \cite{bottou2010large}. Since an exhaustive survey of all these possibilities is beyond the scope of this letter, we just limit the discussion to the simplest setting, which may then be seen as a starting point for improvements.

\section{Example: scalar field in three dimensions}
We illustrate the working of our general algorithm based on a scalar field theory in a three-dimensional Euclidean spacetime. The RG flow possesses a NGFP, the Wilson-Fisher fixed point. We stress that in our example, the Wilson-Fisher fixed point does not appear in its usual role as an IR-fixed point describing a critical phenomenon after fine-tuning the RG flow to its IR-attractive hypersurface. Instead we build a UV-completion based on this fixed point and determine the predictions resulting from departing from the fixed point along its UV-repulsive eigendirection.

Working with the Wilson-Fisher fixed point comes with the advantage that it is already seen in simple approximations. For our purpose, it suffices to work with the local potential approximation (LPA) \cite{wetterich1993exact}. In this case, the effective average action is approximated by
\be\label{LPA}
\Gamma_k[\phi] \simeq \int d^3x \left[ \frac{1}{2} (\p \phi)^2 + U_k[\phi] \right] \, , 
\ee
where $\phi(x)$ is a real scalar field with mass-dimension zero. Implementing our convention on the couplings, we expand the potential according to
\begin{equation}\label{Upot}
   U_k[\phi] =\frac{1}{4! \, \tilde{u}_k} \phi^{4}+\sum_{0 < n \, , \, n\neq 2}^{N_{\rm max}}\frac{v^n_k}{(2n)!} \, (\tilde{u}_k)^{n-3} \, (\phi^{2})^n \, .
\end{equation}

This singles out the coupling $\tilde{u}_k$ (taken with a negative mass dimension) associated with the $\phi^4$-term as the free parameter allowed by asymptotic safety. The IR-value of the dimensionless couplings $v^n = (v^1, v^3, v^4,\cdots)$ will be predicted based on the UV-critical surface of the NGFP.

\begin{table*}[t]
  \centering
  \begin{tabular}{| c || c | c | c | c | c || c | c | c | c | c || c | c | c | c |}
    \hline
   $\Big. N_{\rm max}$ & $u_*$ & $v^1_*$ & $v^3_*$ & $v^4_*$ & $v^5_*$ & $\theta_1$ & $\tilde{\theta}_1$ & $\tilde{\theta}_2$ & $\tilde{\theta}_3$ & $\tilde{\theta}_4$& $v^1$  & $v^3$ & $v^4$ & $v^5$\\
    \hline
    $\Big. 4$ & $0.129$ &  $−0.0013$  & $ $ & $ $ & $ $ & $1.84$ & $−1.18$  & $ $ & $ $ & $ $ & $0.0048$& $ $ & $ $ & $ $ \\
    \hline 
    $\Big. 6$ & $0.081$ & $−0.00092$  & $541$ & $ $ & $ $& $1.69$ & $ -1.13$ &  $-12.20$ & $ $ & $ $& $0.0017$ & $1043$ & $ $ & $ $ \\ 
    \hline 
    $\Big. 8$ & $0.071$ & $-0.00088$  & $889$ & $1167462$ & $ $& $1.58$ & $-0.96$ &  $-8.82$ & $-32.52$ & $ $& $0.0013$  & $1581$ & $1652938$ & $ $  \\ 
    \hline 
    $\Big. 10$ & $ 0.068$ & $-0.00087$  & $1021$ & $1786707$ & $3427010600$& $ 1.54$ & $-0.78$ &  $-6.52$ & $-22.00$ & $-60.30$& $0.0012$ & $1738$ & $2169522$ & $-3615823144$ \\ 
    \hline
  \end{tabular}
  \caption{Properties of the Wilson-Fisher FP obtained from the LPA, eq.\ \eqref{Upot}, up to order $N_{\rm max} = 10$. The first and second block give the position and stability coefficients of the fixed point, respectively. The values $v^\mu \equiv \lim_{k\rightarrow 0} v_k^\mu$ predicted by asymptotic safety are summarized in the third block.}
  \label{tab:1}
\end{table*}
In Table \ref{tab:1} we report results up to order $N_{\rm max} = 10$.\footnote{These results serve as an illustration of the method and should not be mistaken as an attempt of a precision computation. The latter may be achieved by extending the approximation along the lines \cite{Balog:2019rrg,Delamotte:2024xhn}.} For simplicity, we limit our exposition to the simplest case where $N_{\rm max} = 2$, i.e., we are dealing with two couplings $(u,v)$ and refer to the supplementary notebook for the detailed implementation of this case.

We start from the beta functions given in \cite{reuter2019quantum}. Adapting to our parameterization of the potential and setting $N_{\rm max} = 2$ and $d=3$, we have 
\begin{equation}\label{betaexample}
\begin{split}
    \beta_{u} = & \, u-\frac{u{}^6}{\pi^2  \left(u^2 + v\right){}^3}, \\
    \beta_{v}= & \, -\frac{u{}^5}{6 \pi^2  \left(u^2+v\right){}^2} - \frac{2 \, v \, u^5}{\pi^2 \left( u^2 + v \right)^3} \, .
\end{split}
\end{equation}
The first term in $\beta_v$ is associated with the scale-dependence of $v$. The second term is proportional to the quantum part of $\beta_u$ and captures the contribution of an anomalous scaling dimension associated with $\tilde{u}_k$.

\begin{figure}[ht]
    \centering
    \includegraphics[width=\columnwidth]{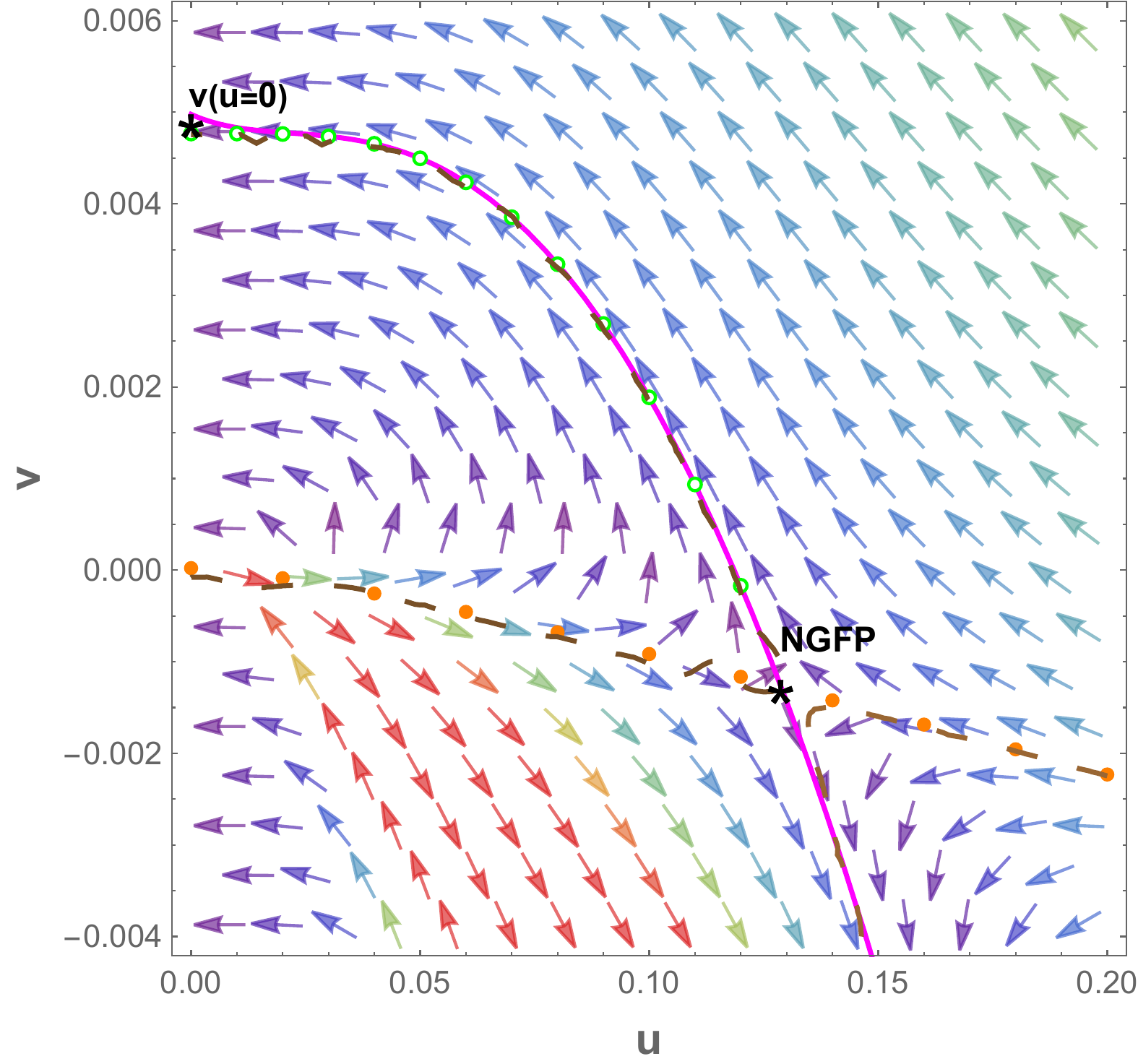}
    \caption{\label{fig:2} Phase diagram generated by the beta functions \eqref{betaexample} with arrows pointing towards lower values of $k$. The NGFP and the endpoint of $\cS_{\rm UV}$ giving rise to the prediction \eqref{prediction-final-ultimo-finito} are marked by $*$. The UV-critical surface obtained from the shooting method is marked by hollow green dots, the solution of the implicit method by brown dashes, and the one of the explicit method by a solid magenta line. For completeness we added the RG trajectories generated by $\tilde{V}_1$ as solid orange dots.}
\end{figure}

The NGFP of the system is located at $(u_*,v_*) = (\frac{2197}{1728 \pi^2}, - \frac{371293}{2985984 \pi^4})$. The stability coefficients and right-eigenvectors entering into the linearized solution \eqref{linsol} are
\be
\begin{split}
\theta_1 = \frac{1}{6}(2+\sqrt{82}) \, , \; & 
V_1 = \left(\frac{864 \left(\sqrt{82}-11\right) \pi ^2}{2197} \, , \, 1 \right)^{\rm T} \, ,  \\
\tilde{\theta}_1 = \frac{1}{6}(2-\sqrt{82})\, , \; & 
\tilde{V}_1 = \left(-\frac{864 \left(\sqrt{82}+11\right) \pi ^2}{2197}\, , \, 1 \right)^{\rm T} \, .
\end{split}
\ee
Thus we are dealing with a saddle-point with one UV-attractive and one UV-repulsive eigendirection. The UV-critical surface is one-dimensional and we expect a prediction for $v$ from asymptotic safety. Following the procedure leading to \eqref{surface2}, the generating function for the UV-critical surface  is
\be
\begin{split}
 F^\ast(u,v) =
-u+\frac{864 \left(\sqrt{82}-11\right) \pi ^2 v}{2197}+\frac{169 \left(15+\sqrt{82}\right)}{3456 \pi ^2} \, ,
\end{split}
\ee
where we have chosen the normalization of the normal vector $n_\mu$ such that the coefficient multiplying $u$ is set to one. Based on this input we compute the prediction for $v$ using 1) the traditional shooting method,
2) a solution of the explicit system \eqref{master2} based on series expansions of orders 10 to 15, and 3) a spectral solution evaluating the implicit equation \eqref{master1} for a linear combination of 500 MCDs on a regular lattice centered on the NGFP. All methods yield
\be\label{prediction-final-ultimo-finito}
v \equiv \lim_{k \rightarrow 0} v_k \simeq 4.8 \times 10^{-3} \, , 
\ee
with the numerical difference appearing in the third relevant digit. Fig.\ \ref{fig:2} illustrates the structures underlying this prediction.

\section{Summary and Outlook}
The Wilsonian renormalization group (RG) is a powerful tool to study the effect of fluctuations in statistical and quantum systems. Typically, such investigations proceed along the following steps. 1) one computes the beta functions of the theory in a suitable approximation. On this basis, one determines the RG fixed points and their stability properties. 2) In the case of asymptotic safety, where the UV-completion is provided by an interacting fixed point, one constructs the UV-critical surface of the fixed point in order to obtain the set of effective actions compatible with this UV-completion. 3) The output of step 2) is used to construct physical observables which allow to confront the predictions with observations. 

Our work proposes an efficient algorithm for carrying out the second step in this procedure, by solving multi-linear partial differential equations for the generating functions encoding the embedding of the UV-critical surface. This constitutes a significant improvement compared to the usual shooting method  (employed e.g., in \cite{Morris:1996xq,Bervillier:2007tc,Reuter:2003ca,Shaposhnikov:2012zz,Eichhorn:2017ylw,Gubitosi:2018gsl,Basile:2021krr,Baldazzi:2023pep}) which traces individual RG trajectories. We illustrated the working of our algorithm based on the Wilson-Fisher fixed point. Interpreting this fixed point as the UV-completion of the RG flow, we predicted the couplings of the effective potential based on the single free parameter allowed by asymptotic safety (see Table \ref{tab:1}). 

We stress that the methods described in this letter are applicable in a much broader context. Potential applications beyond the scope of the present work are the gravitational asymptotic safety program \cite{percacci2017introduction,reuter2019quantum},  including its extension by  matter degrees of freedom \cite{Eichhorn:2022gku,Pastor-Gutierrez:2022nki}, asymptotically safe gauge theories \cite{Litim:2014uca,Bond:2017wut,Hiller:2020fbu,Bond:2022xvr,Litim:2023tym} and their supersymmetric extensions \cite{Intriligator:2015xxa,Bajc:2016efj,Bajc:2017xwx,Bond:2017suy,Hiller:2022hgt}, and asymptotically safe quantum electrodynamics \cite{Gies:2020xuh,Gies:2022aiz,Gies:2024ugz}. The authors have already tested its applicability for asymptotically safe scalar-tensor theories \cite{Narain:2009fy,Eichhorn:2012va,Henz:2013oxa,Percacci:2015wwa,Labus:2015ska,Henz:2016aoh,Becker:2017tcx,Eichhorn:2017als,Pawlowski:2018ixd,Wetterich:2019rsn,Wetterich:2019zdo,Eichhorn:2020sbo,Ohta:2021bkc,Laporte:2021kyp,Laporte:2022ziz,Knorr:2022ilz,Wetterich:2022ncl,deBrito:2023myf} and the resulting shapes of the effective potentials will be reported in \cite{Silva:2024wit}. 

\section*{Acknowledgements} We thank M.\ Becker, C.\ Laporte, R.\ Loll, and J.\ Wang for useful discussions during the process of developing these ideas.

\bibliography{bibliography}

\end{document}